\documentclass[12pt]{article}
\usepackage{graphicx,url}
\usepackage{amsmath,amssymb,amsfonts,cite}
\usepackage[paper=letterpaper,margin=0.9in]{geometry}

\newcommand{\be}{\begin{equation}}
\newcommand{\ee}{\end{equation}}
\newcommand{\bea}{\begin{eqnarray}}
\newcommand{\eea}{\end{eqnarray}}
\newcommand{\ba}{\begin{array}}
\newcommand{\ea}{\end{array}}

\newcommand{\comment}[1]{}

\def\ffract#1#2{\raise .3 em\hbox{$\scriptstyle#1$}\kern-.25em/
                \kern-.2em\lower .2 em \hbox{$\scriptstyle#2$}}

\def\part#1#2{{\partial#1\over\partial#2}}

\begin{document}

\title{Probing the Climatological Impact of a Cosmic Ray-Cloud Connection through Low-Frequency Radio Observations}
\date{}

\renewcommand{\thepage}{\arabic{page}}
\setcounter{page}{1}


\vskip 2.00 cm
\renewcommand{\thefootnote}{\fnsymbol{footnote}}
\maketitle

\comment{
\centerline{\bf \Large Probing the Climatological Impact of a Cosmic Ray-Cloud} 
\centerline{\bf \Large connection through low frequency radio observations}}
\vskip 0.75 cm

\centerline{{\bf Nathan Magee${}^{1}\footnote{\tt magee@tcnj.edu} \&$
Michael Kavic,${}^{1}$\footnote{\tt kavicm@tcnj.edu}}}

\vskip .5cm
\centerline{${}^1$\it Department of Physics}
\centerline{\it The College of New Jersey}
\centerline{\it 2000 Pennington Rd.}
\centerline{\it Ewing, NJ 08628, U.S.A.}
\vskip .5cm
\setcounter{footnote}{0}
\renewcommand{\thefootnote}{\arabic{footnote}}

\begin{abstract}
It has been proposed that cosmic ray events could have a causal relationship with cloud formation rates. Given the weak constraints on the role that cloud formation plays in climate forcing it is essential to understand the role such a relationship could have in shaping the Earth's climate. This issue has been previously investigated in the context of the long-term effect of cosmic ray events on climate. However, in order to establish whether or not such a relationship exists, measurements of short-timescale solar events, individual cosmic ray events, and spatially correlated cloud parameters could be of great significance. Here we propose such a comparison using observations from a pair of radio telescopes arrays, the Long Wavelength Array (LWA) and the Eight-meter-wavelength Transient Array (ETA). These low-frequency radio arrays have a unique ability to simultaneously conduct solar, ionospheric and cosmic rays observations and are thus ideal for such a comparison. We will outline plans for a comparison using data from these instruments, satellite images of cloud formation as well as expected cloud formation rates from numerical models. We present some preliminary results illustrating the efficacy of this type of comparison and discuss future plans to carryout this program.
\end{abstract}


\newpage

\section{Introduction} 
Cloud formation processes and associated climate feedbacks are among the most poorly understood aspects of climate forcing \cite{Solomon09, Hegg09}. The proposal that cosmic rays could play a significant role in cloud formation due to the increased ionization engendered by such events was first proposed by Wilson at the beginning of the last century. \cite{Wilson}. This issue has been considered in the context of variations in solar activity \cite{FRIIS-CHRISTENSEN01111991}. It has been suggested that modulation of galactic cosmic rays (GCRs) by the solar wind could lead to reduced cloud formation \cite{2002GeoRL..29w..22K}. A demonstrable reduction in GRC events is routinely observed during Earth-bound Coronal Mass Ejections (CMEs)\cite{Wawrzynczak2005682}, the so-called Forbush effect. 

A potential connection between GCR flux and cloud properties has been analyzed in the context of long time periods (i.e. months or years) \cite{Svensmark97, Carslaw02, Laut03, Svensmark09, Rohs10} and in a controlled experimental environment \cite{springerlink:10.1007/s10712-008-9030-6}.  The climate-scale data analyses have often, though not always, shown some evidence of correlations between GCR fluxes and cloudiness fractions.  In the instances where correlations have been shown, the purported connections have usually been highly qualified on the bases of geography, time-span, cloud-type, and/or have been the subject of considerable debate over analyses methods and statistical significance.  Several of the  recent analyses have focused on shorter time-scale correlations (with resolution of fractions of days) between GCR flux and cloud properties \cite{Brown2008, Rohs10}. While these studies have examined possible correlations and lags on a global scale, the assimilated horizontal cloud data resolution has still been very coarse (min. of 30 km) compared to real cloud-scale processes.  Furthermore, in most instances, a single, non-spatially distributed, GCR proxy has been used as the correlation basis.  There has also been one recent effort to observe potential effects on nucleation and cloud formation due to a strong solar event linked with solar energetic particles (SEPs) \cite{2008GeoRL..3518610M}. In the context of these contending analyses, we describe a new multi-scale and multi-modal approach to this question using several streams of high temporally and spatially-resolved data.  Novel solar, ionospheric, and cosmic ray data will be acquired using two low frequency transient radio arrays, the Long Wavelength Array (LWA) and the Eight-meter-wavelength Transient Array (ETA).  These data will be supplemented by 1) spatially-resolved ionospheric measurements derived from GPS receivers, 2) a global network of neutron monitors of local GCR flux, and 3) a suite of geomagnetic and solar indices. The supplemental data streams are already available in near-real time and each is characterized by high spatial and temporal resolutions (described specifically in section 2.5).

There are two predominate mechanisms by which cosmic rays and space weather have been hypothesized to influence cloud properties. The first, and most widely-tested, mechanism involves increased aerosol production due to enhanced ionization caused by cosmic ray-induced particle cascades in the troposphere \cite{springerlink:10.1007/s10712-008-9030-6}. Cosmic rays collide with elements of the Earth's atmosphere, typically oxygen or nitrogen molecules, and generate ionized molecules. These ionized molecules may enhance the ability of aerosols to form Cloud Condensation Nuclei (CCN) by increasing binding of small proto-CCN clusters of molecules, allowing for growth beyond a critical size past which condensation becomes more likely. The second mechanism involves cosmic and solar modulation of the Global Electric Circuit (GEC) \cite{Siingh07, Harrison10, springerlink:10.1007/s10712-008-9030-6}. The GEC is characterized by a fair-weather electric field of ~100 V/m near the surface and complex and dynamic patterns of electrical currents \cite{Tinsley00, Markson07, Mach09}. The GEC is generated by an electric potential (global mean, 240 kV) which is maintained between the ionosphere and the ground largely through continual lightning strikes. Short and long-scale variability in the GEC has been hypothesized to effect cloud processes through enhancement of ionization-related aerosol scavenging  or through modulation of cloud dynamical/microphysical feedbacks \cite{Tinsley08, Rycroft00, springerlink:10.1007/s10712-004-5439-8}.

\section{Investigation the Cosmic Ray-Cloud Connection with Low-Frequency Radio Telescope Arrays}

Low-frequency radio telescope arrays are uniquely poised to investigate the possible existences of a cosmic ray-cloud connection. This type of instrument can provide observational data on a surprising number of different aspects of this phenomena. They are able to observe solar radio bursts and the initial creation of a CME and can track CMEs as they move through interplanetary space. They are also capable of detailed real-time tomographic ionospheric mapping, including ionospheric disturbances caused by solar storms and solar cosmic rays, and low-frequency radio arrays can be used to directly observe the coherent radio signal emitted during a cosmic ray event. 
 
The LWA and ETA are two radio telescope arrays capable of conducting the types of observations outlined above. The LWA is composed of dual polarization, active antennas sensitive to a bandwidth of 10-88 MHz. It is organized into stations which are themselves composed of 256 antennas each. The first station, LWA-1, has been completed near Socorro, NM and has recently begun conducting observations. Each dipole antenna is direct-sampled (196 MHz A/D). This allows for the creation of four beams which can independently scan different portions of the sky and the available band-width. These beams are rapidly reconfigurable. When configured for atmospheric/solar observations, one of the four beams will track the sun while a second beam provides near real-time, full-sky ionospheric monitoring \cite{2010amos.confE..59K,Ellingson,LWAold}. The ETA is a low-frequency radio array located in western Virginia. ETA's instrumentation is similar to that of LWA-1 and will conduct coincident observations with the LWA. The ETA is composed of 10-dual polarization antennas which are sensitive to a frequency range of $29-49$ MHz \cite{ETA1,ETA2}. 

\begin{figure}[h!]
\begin{center}
\includegraphics[width=100mm]{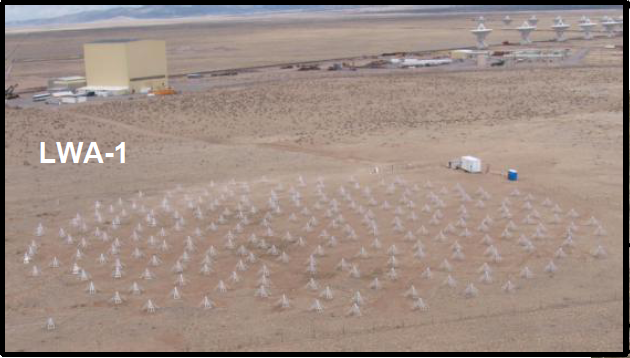}
\caption{The first station of the Long Wavelength Array (LWA-1) located in Socorro, NM. The LWA-1 station is $\sim100$m in diameter. }
\label{fig:cube2}
\end{center}
\end{figure}

\subsection{Solar observations}
Given the steerable multi-beam, multi-frequency capability of the LWA/ETA, dedicated solar monitoring is possible. This allows for detection of solar burst events whose associated radio emission is strongest within the bandwidth to which LWA/ETA is sensitive. The radio emission from a solar burst is caused by the synchrotron emission and in the case of a CME from electrons accelerated in the CME-driven shock above the solar corona. This has been observed with Nancay Radioheliograph at 164 MHz by \cite{2001ApJ...558L..65B}. This type of emission are nominally weaker that coherent plasma emission, but imaging reveals a quasicircular arc of emission co-located with the CME front. The spectrum of this emission close to the Sun \cite{2001ApJ...558L..65B} appears to be consistent with a nonthermal spectrum suppressed at low frequencies by the Razin effect. The LWA/ETA will be able to trace such emission to much greater heights above the Sun due to its lower frequency range. Razin suppression will be less effective at greater heights due to the lower densities in the atmosphere there, and the emission should be very bright and easily observed in the LWA/ETA frequency range.

\begin{figure}[h!]
\begin{center}
\includegraphics[width=180mm]{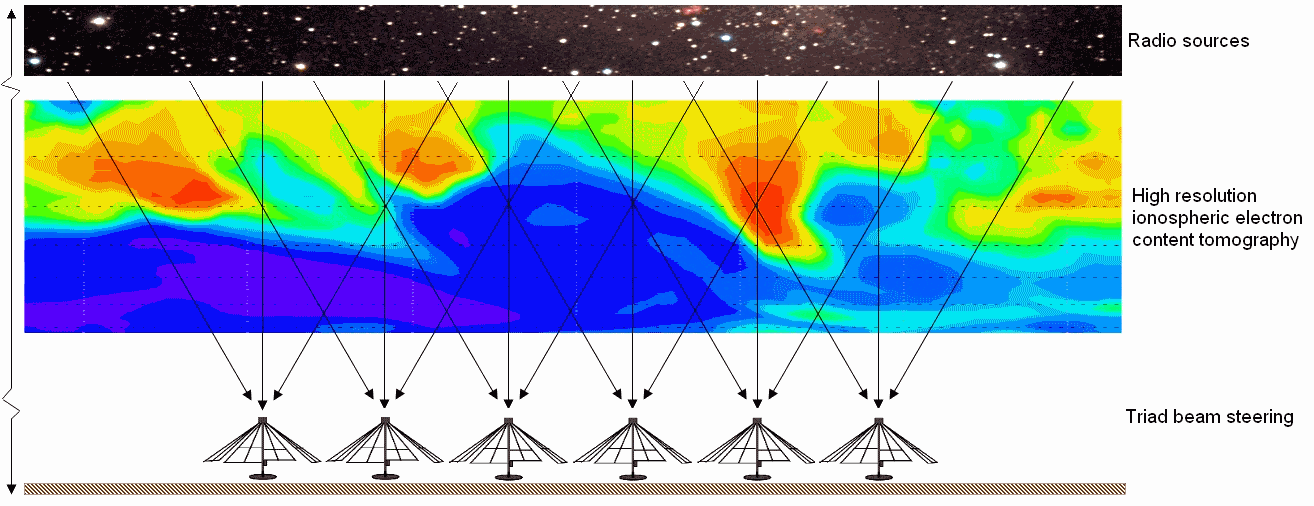} 
\caption{A variation in the phase of a radio signal is induced as it transverses the ionosphere. Using observations from an arrays of radio telescopes such as LWA/ETA/ASTRA.}
\label{fig1}
\end{center}
\end{figure}

\subsection{Interplanetary Tracking of CMEs}

LWA/ETA is able to track CMEs using a technique known as interplanetary scintillation (IPS). This method involves tracking regions of increased density of the solar wind associated with CMEs as they propagate through the interplanetary medium. This approach has been used successfully for similar observation by several other groups \cite{springerlink:10.1007/BF00670233}. Scintillation can be described as fluctuation in a radio signal due to an intervening medium. In the case of IPS, the signal from bright radio sources are diffracted and scattered by density fluctuation in the density of the solar wind. Moreover, the degree to which a radio signal is scattered is proportional to the absolute density of the solar wind. Thus a CME would produce a high degree of scintillation which allows it to be tracked with a radio telescope \cite{springerlink:10.1007/s11207-009-9350-9}. This has been done with so called transit instruments. However, such instruments can only observe a bright radio source daily as it crosses the local meridian and are thereby limited in the ability to track such events. The flexible beam-forming capabilities of the LWA/ETA do not require mechanical adjustment of the instrument as beams are formed electronically through phasing of groups of antennas. Thus multiple bright radio sources can be sampled simultaneously with millisecond time resolution and new sources can quickly be acquire to flow the trajectory of the CME. Moreover, the wide frequency range over which the LWA can observe allows for use of a frequency-correlation technique to obtain velocity measurements. Using this technique the first station LWA-1 and ETA can be used to track CMEs. In principle, the early identification and tracking of CMEs could by used to trigger spatially and temporally targeted radio observations of the ionosphere.   

\subsection{Ionospheric Observations}

An analogous scintillation effect to the one involved in IPS occurs as the signal from a strong low-frequency radio source transverses the Earth's atmosphere. A change in the refractive index is caused by the distribution of electrons in the Earth's ionosphere. This in turn changes the phase velocity and introduces a path-dependent delay. 
This phase shift is directly related to the TEC
\be
\phi=\frac{e^2}{cm_e\nu}\int N_e(l)dl,
\ee
where $\nu$ is the frequency of the radio signal and $N_e(l)$ is the electron column density.
Thus a radio array composed of multiple antennas can be used for extremely accurate measurements of small variations in electron density and total electron content (TEC). Standard ionospheric observing techniques are sensitive to variation in TEC ($\Delta$ TEC) at the .1 TECU level (1 TECU$=10^{12}$ electrons per cm$^{2}$) and can detect spatial variations on a scale of $\sim100$km. Radio telescopes can improve on these limits; the Very Large Array (VLA) for example could be used to reach milli-TECU sensitivities and spatial variations with resolutions of $\sim 100$m. Ionospheric observation by the LWA/ETA will significantly improve sensitivity even as compared to these VLA observations, potentially to the $\sim 10$m scale.  LWA/ETA are instruments designed to be dynamically sensitive to wide low-frequency bandwidth and multiple, electronically steered beams. This low-frequency spectrum is prohibitive in a certain sense for some astronomical observations because of ionospheric interference. However, this presents a significant opportunity for ionospheric observations. A single station of the LWA (LWA-1) can study the ionosphere by rapidly cycling through observation of strong, compact radio sources in various parts of the sky and measuring the delay along each line of sight as in Fig. \ref{fig1}. Thus ionospheric events such as sudden ionospheric disturbances (SIDs) will appear in LWA data as a spatially varying phase-shift along each line of sight. The radio burst from a solar flare inception can be observed as noted above. When the X-rays from the flare reach the ionosphere, they will ionize the D-region, making radio propagation more difficult. This is observable in the LWA/ETA. Solar flares and the shock-wave associated with CMEs also emit solar energetic particles (SEPs) which in the case of solar flares arrive at Earth $\sim20$ mins after the radiation. These particles ionize the D and E regions of the ionosphere which could also be observed by the LWA/ETA. 

\subsection{Cosmic Ray Observations}

When cosmic rays enter the Earth's atmosphere, they collide with molecules, typically oxygen or nitrogen creating extensive air showers (EAS) of lighter charged particles. These particles in turn generate synchrotron radiation as they move in the Earth's magnetic field. This radiation is emitted coherently up to $\sim100$MHz. This coherent low frequency radiation is detectable in low frequency radio arrays including LWA/ETA. The LOPES (Lofar Prototype Station) experiment has demonstrated the viability of detecting cosmic ray air showers using low frequency radio signals. \cite{2003ICRC....2..969H, LOPES} We propose that detection of Galactic Cosmic Ray (GCR) incidence can be similarly detected and described by the LWA.  It has been established that the incidence of GCR particles is not isotropic, even when integrated over long time scales \cite{OyamaSuperK}. Over short time scales, this spatial anisotropy is expected to be even more pronounced and should be used directly when considering hypothesized  impacts on cloud processes. The use of the LWA/ETA and similar facilities will allow for assessment of ionization-induced cloud effects that might occur on timescale of hours to days and with pronounced spatial variability. 
 
\subsection{Other Source of Solar, Ionospheric, \& Cosmic Ray Data}

While we anticipate that low frequency radio observations will provide an novel resolution benchmark in ionospheric imaging, it is also useful to employ a broad suite of available geophysical data that might be mechanistically tied to a cloud/climate connection.  In several analyses, cosmic ray fluxes as detected through neutron monitoring at one site (Moscow) are central to assertions of cloud/climate correlations \cite{Brown2008, Svensmark09, Krist08}.  Cosmic ray fluxes have been hypothesized to modulate aerosol activation and cloud particle nucleation events, so it is readily apparent that cosmic ray data should be considered.  However, several other relevant geomagnetic and geoelectrical parameters are also routinely measured with high resolution by ionosonde, satellite sensors, and GPS receivers.    

In addition to the widely-used Moscow Neutron Monitor, cosmic ray flux is also available from a broadly distributed set of seven Neutron Monitoring stations operated by the Bartol Research Institute at the University of Delaware \cite{Neut}.  Archives of neutron counts from these stations are available at 1 hours intervals. The currently operational monitoring stations are located at:  McMurdo, Antarctica; South Pole, Antarctica; Thule, Greenland; Newark, Delaware; Inuvik, North West Territories; Fort Smith, North West Territories; and Nain, Labrador.  Neutron data is also available from approximately 15 other globally distributed monitoring stations that are independently operated by universities and observatories.  While cosmic flux data recorded at disparate locations show very strong correlations, they are not identical, reflecting anisotropic cosmic sources and effects of geospatial and secular modulation.  For disparate regions, the timing and magnitude of cascade products reaching to the surface can vary significantly, and could be expected to affect cloud processes differently.  
  
Ionospheric total electron content in vertical and slant directions has been determined through processing of GPS transmissions since the late 1990s  \cite{Manucci98}.  Archival data from the National Geophysical Data Center (NGDC) \cite{Tec} is available at 15 minute increments and with spatial resolution of 1 degree latitude and longitude.  These data presently represent the highest spatial and temporal resolution observations of the ionosphere and are therefore an important benchmark for correlation with cloud-resolving parameters, especially in evaluating potential impacts of GEC variability on cloud microphysics.           

In addition, archived collections of global geomagnetic indices are available from the NGDC \cite{Gonzalez94, GIB}. Many of the indices are calculated by International Service of Geomagnetic Indices, Institut de Physique du Globe de Paris, integrating multiple data sources.  Indices are available at 3 hour intervals and include Kp Planetary Index, Equivalent Amplitude A Index, Disturbance Amplitude-Storm Time, Bartel Cycle day, magnetically quiet and disturbed days monthly rank, and aa index.  An analogous bulletin is also available from the NGDC \cite{SIB} for solar activity indices including sunspot number and solar radiofrequency fluxes.

\section{Comparison of cloud cover data with LWA/ETA and other geophysical observations}

Svensmark's approaches \cite{Svensmark97, Svensmark09} to time series correlations have garnered much attention and controversy.  They largely examine time-series correlations in GCR ground-based proxies from neutron monitors and cloud fractions from the International Cloud Climatology (ISCCP) project at monthly temporal resolution.  Brown \cite{Brown2008} examines much of the same data, but at shorter time-scales (3 hour resolution). They find correlations at long and short time scales, but subsequent tests of the same data \cite{Krist08} have raised questions about analysis methods and failed to establish a compelling case for broad statistical significance. The comparison of GCR flux and measured cloud amount clearly has emerged as an important and controversial metric, however the interpretation of times series of raw cloud amount are vastly complicated by evolving weather patterns and natural cloud variability.  To help address these difficulties, we would like to extend cloud comparisons on several fronts: 1) use of A-Train satellite data for a targeted examination of event-specific cloud-microphysics, including densities and size-spectra of aerosol populations, 2) use of operational Global Forecast System numerical weather model (GFS) prognostic values for cloud parameters, considered alongside satellite cloud data, and 3) expanded assimilation of GOES cloud properties.  The simultaneous refinement of cloud and GCR/GEC analyses should offer a newly detailed picture of the potential for a climatological role in the multi-scale connectivity of the atmosphere-near space environment. Resolutions of presently available cloud measurements exceed the spatial resolutions of all of the supplementary data sources described in section 2.5 above, which highlights the critical importance the new LWA-ETA ionospheric imaging capacity; only when the cloud and cosmic regimes are sampled at comparable (high) resolutions, will the full power of multi-stream comparative analyses be accessible. 

In addition to extended sensitivity to anomalous cloud formation via GFS model comparisons and expanded GOES analyses, we also suggest that cloud analysis can be extended beyond the passive radiance sensitivities of goestationary satellites. New suites of cloud and aerosol data have recently become available with the polar-orbiting CloudSat and CALIPSO NASA missions that include active downward-pointing LIDAR and RADAR systems. These data include newly detailed pictures of aerosol properties, multi-layer cloud categorizations, and analyses of total cloud optical depth and cloud water and ice content. When this new data is added to that already available from MODIS and MISR, newly comprehensive and detailed pictures of clouds are now accessible. We intend to use the following data products for comparison to LWA, GCR, GEC, GOES, and GFS data: 
\begin{itemize}
\item
MISR: Multi-angle Imaging SpectroRadiometer
	Cloud Fraction by Altitude (CFbA)
\item
MODIS: Moderate-resolution Imaging Spectroradiometer
	High Resolution Visible Imaging (250m)	
	Cloud Optical Thickness
	Cloud Effective Radius
	Aerosol Optical Depth
	Cloud Fraction (Daytime)
\item
CALIPSO: Cloud-Aerosol Lidar and Infrared Pathfinder Satellite Observation 
	0.5-30 km height LIDAR Aerosol Backscatter Profile 						
\item
CloudSat:	
	Cloud Optical Depth	
	Cloud Water Content (Radar-Visible Optical Depth) 
\end{itemize}
The 2nd generation Cloud Feedback Model Intercomparison Project (CFMIP) is a large international collaboration presently underway, with the scientific goal of establishing comparative quantitative results of cloud feedback effects in more than 20 of the most prominent atmospheric general circulation models. This effort is a response to a charge from the International Panel on Clouds and Climate (IPCC) and is expected to inform the scientific findings for dissemination in the 5th Assessment Report (AR5), working group I, due in fall 2013.  Notably, the AR5 reference document (http://www.ipcc.ch/pdf/ar5/ar5-outline-compilation.pdf) also calls for working group I to compile a review of scientific evidence on the coupling of cloud processes and cosmic rays. To prepare for this process the PFMIP  has produced the CFMIP Observation Simulator Packge (COSP) in order to simplify comparison of A-Train data and ISSCP data with numerical models. The COSP system produces output that matches what CloudSat/CALIPSO/ISSCP would see in a model-generated world, allowing for direct comparisons. It is a flexible tool to simulate active instruments in all types of atmospheric models (climate, forecast, cloud-resolving) and can be directly applied as a new and sophisticated numerical tool for model/satellite comparisons.  While this approach clearly portends important discoveries, it is also important to note that polar-orbiting satellites are inherently limited in tracking time-resolved processes:  A-Train sensors including MODIS,CALIPSO,and CloudSat make overpasses of the same location only twice daily.  For this reason, geostationary satellite products offer critical measurements for analyses of cloud processes at the time scales that are inherent to individual cloud systems (on the order of minutes to hours to days). 

The cloud scheme \cite{Xu96} within the GFS model makes predictive estimates of cloud fraction and cloud water at all model grid points with a 27km global spatial resolution and at 3 hour increments \cite{GFSvalidation, Ye10}. Cloud fraction is calculated at all 64 vertical model levels and can be broken down into high, middle, and low cloud bins for the sake of comparison with satellite-derived values.  Other forecast parameters of potential interest include column-total cloud water content and relative humidity at all model levels.  The most accurate forecasts would be (arguably) taken from the GFS model initiated (00z and 12z) most proximately prior to the time of interest. This comparative approach would allow the satellite measurements to be compared against the expected cloud amounts. This should permit a more sensitive identification of instances of unexpectedly low or unexpectedly high cloud formation, rather than relying solely on raw cloud measurements which are inherently difficult to separate from complex weather patterns. To the extent that anomalous cloud events are associated, or not associated, with GCR variations, solar energetic particle events, and GEC modulations, it should be possible to expand the scientific confidence in the significance or absence of any cosmic/cloud connection.  The use of high spatial and temporal resolutions for cloud, GCR, and ionospheric parameters would represent a marked departure from previous studies.

Furthermore, goestationary satellite-derived ISSCP cloud data products have been most widely used for correlation analyses.  While these data are incontrovertibly excellent sources for climate-scale comparisons, they also have major limitations in that data products are not rapidly available (currently only processed into mid-2008) and that they are available at drastically reduced spatial and temporal resolutions.  Goestationary satellite measurements are available for all locations over the continental US at approximately 15 minute intervals and with 1 km resolution at visible wavelengths and 4 km resolution in 4 infrared channels.  By contrast, the finest resolution ISSCP data is available at 30 km spatial resolution and 3 hour time increments.  Furthermore, many of the most relevant imaging and sounding measurements of the GOES sensors are not available in the ISSCP products.  For these, reasons, we directly access GOES data streams at highest available resolution and including derived parameters such as cloud-top temperatures, cloud-top pressure, cloud optical depths.

\begin{figure}[h!]
\begin{center}
\includegraphics[width=170mm]{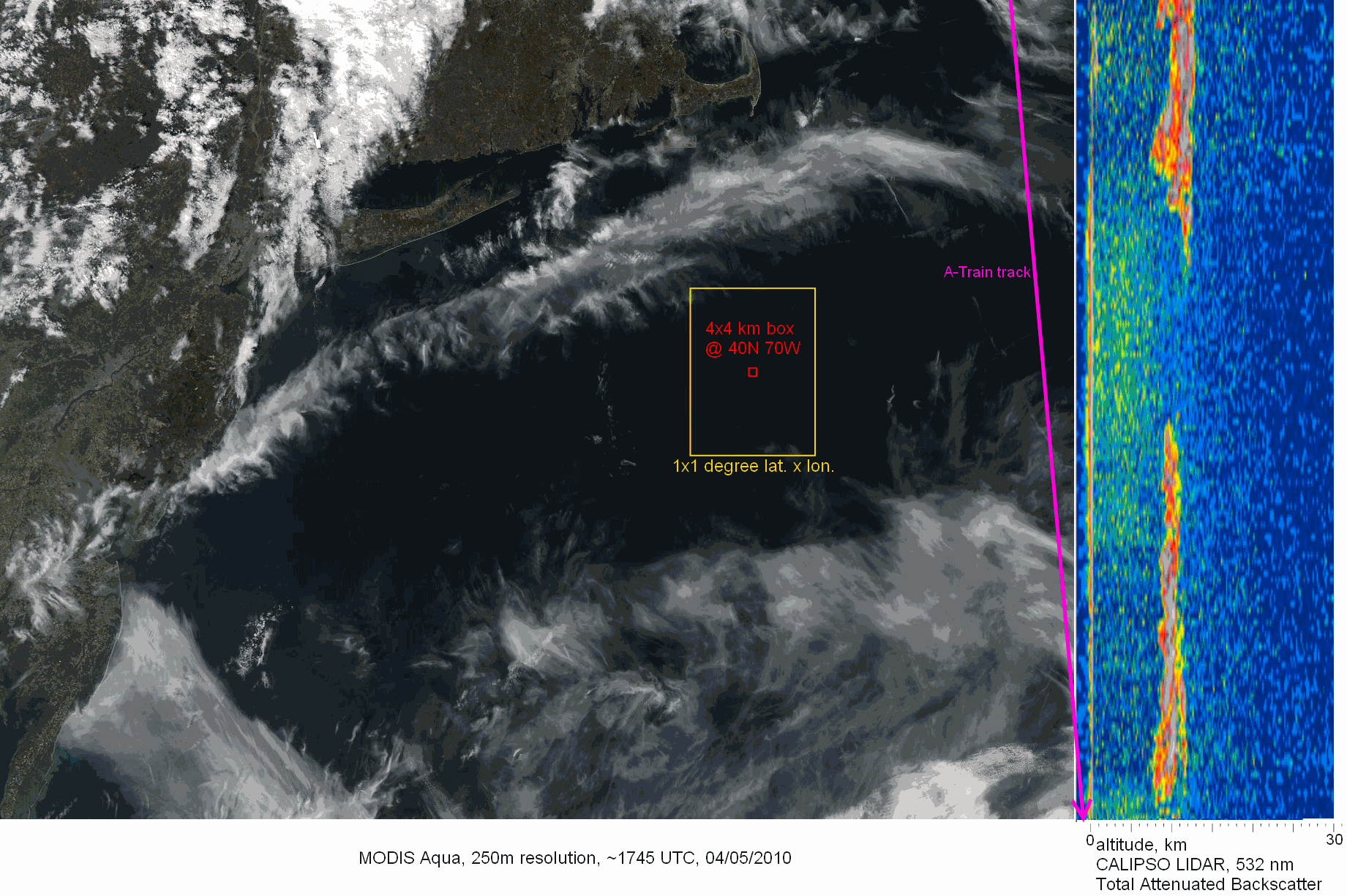}
\caption{MODIS aqua, 1745 UTC, 4/5/2010 true-color image of 40N,70W benchmark, 250m resolution. Yellow rectangle bounds a 1x1 degree latitude x longitude box, corresponding to the resolution of TEC data.  The red rectangle is a 4x4 km box, corresponding to the resolution of the infrared GOES-12 satellite bands. At right, a CALIPSO lidar profile (near-simultaneous with MODIS image, taken along pink slice) gives the 532 nm total attenuated backscatter from 0.5 to 30 km}
\label{fig:cube2}
\end{center}
\end{figure}

\section{Sample study of solar event of 04/05/2010}

A confined geographical sample set of satellite, cosmic, geomagnetic, and geoelectrical time series have been retrieved for the most dramatic solar event of 2010, which occurred on April 5.  While April 5, 2010 was the most active geomagnetic day of 2010, the year was quiet and the magnitude of disturbances on this day would occur frequently during periods of peak of solar activity. The purpose of this retrieval is to demonstrate the application of broad suites of data over highly resolved spatial and temporal scales. The region of examination is centered on the location of 40N, 70W.  This location is in an open-ocean environment that is typically north of the gulf stream boundary, approximately 85 miles south of Nantucket and 215 miles east of the New Jersey coast. This specific coordinate set is often referred to as a geographical ``benchmark" by meteorologists because of the proclivity for major east-coast snowstorms to pass near this location.  Despite prevailing offshore winds in this region from the west, the air is usually quite clean by the time it reaches 40N, 70W. Figure 3 displays a 250m resolution image from MODIS-Aqua and a near-simultaneous CALIPSO LIDAR profile taken at approximately 1745 UTC on April 5, 2010. The region directly at 40N, 70W is clear at this time, though a variety of high altitude cirrus are broadly distributed across the region at approximately 11 km altitude. Some stratocumulus are located over land in Connecticut and New York state and some low-altitude marine stratus are in the near-shore Delaware waters. The image also shows some evidence of linear cirrus features (south of Long Island) likely initiated via contrail spreading as well as the appearance of some thin ship-tracks at lower altitude (east and southeast of Cape Cod).  Both of these features suggest an environment where cloud particle nucleation is partly limiting cloud formation processes. The cool maritime environment is characterized by high low-level humidity and sparse aerosol content. The presence of stronger correlations in maritime environments have been suggested by several of the analyses of cosmic/cloud correlations \cite{Krist08, Svensmark09}. If a link between cosmic ray incidence and cloud particle nucleation is present, such a clean maritime environment would be expected to show a more sensitive response than locales where cloud formation would be unlikely to be limited by a dearth of aerosol and cloud condensation nuclei. 

\begin{figure}[h!]
\begin{center}
\includegraphics[width=170mm]{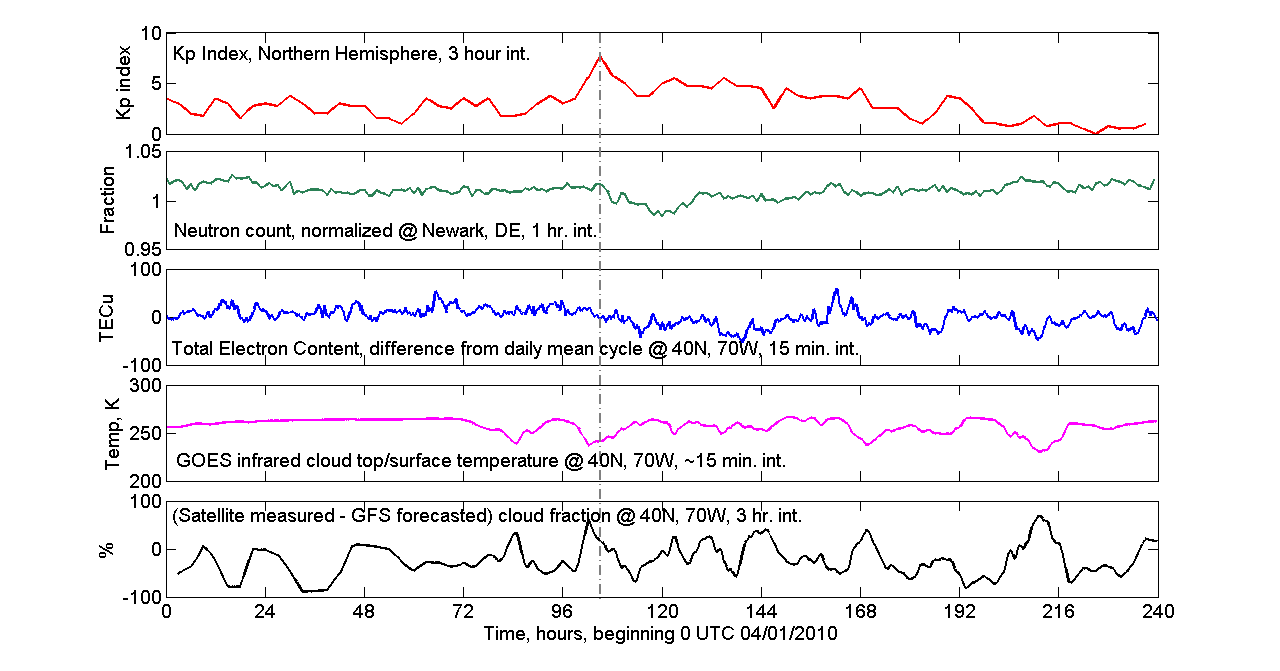} 
\caption{Time series of geophysical data near 40N, 70W  from 0 UTC, 1 April 2010 to 24 UTC, 10 April 2010.  Dashed vertical line indicates beginning of strong solar event associated with a Forbush decrease in GCR flux}
\label{fig:cube2}
\end{center}
\end{figure}

Figure 4 shows a set of 5 times series covering the first 10 days in April, 2010.  All data have been collected at maximum available spatial and temporal resolution (values indicated on Figure 4) and have been re-sampled to give a synchronous set of observations to test for correlations.  As the top panel of northern hemisphere Kp indices indicates, this time period was characterized by modestly active solar and geomagnetic activity during the first 4 days, followed by a strong event arriving at about 8 UTC on the 5th.  The 5th and 6th were the 2 most active days of the month (and year) and the activity declined to the 2nd quietest day of the month on the 10th.  The 2nd panel from top shows the normalized neutron count at Newark, Delaware (about 250 miles west of the benchmark) dropping significantly, precisely in-sync with the Kp index increase.  This marks the beginning of a Forbush decrease in GCRs which is prominent until the end of April 5, and modestly apparent through the 7th of April.  The central panel, in blue, shows the vertical total electron content anomaly in the 1x1 degree region. The 24 hour daily cycle of TEC variation was removed from the raw TEC values in order to reveal anomalously high and low TEC content.  The GOES infrared satellite measurements (derived from channel 6 radiance) show the cloud-top temperature when clouds are present, or show ocean surface/boundary layer temperature when clouds are absent.  The GOES time series corresponds to a largely clear weather pattern from the 1st through the 3rd, followed by intermittent cloudiness, much of which is high-altitude cirrus, as indicated by cloud top temperatures below 250K and through viewing of contemporaneous GOES and MODIS visible satellite imagery. A supplemental movie of IR temperature in a 2 x 2 degree window around the benchmark is available at http://www.tcnj.edu/~magee/cloud lab/  The lowest panel (in black) subtracts the GFS-forecasted cloud fraction from the GOES-derived cloud fraction.  The resulting time series displays periods when measured cloud amount at the benchmark exceeded model predictions (positive values) as well as times when measured clouds were fewer than modeled amounts (negative values).  

\begin{figure}[h!]
\begin{center}
\includegraphics[width=170mm]{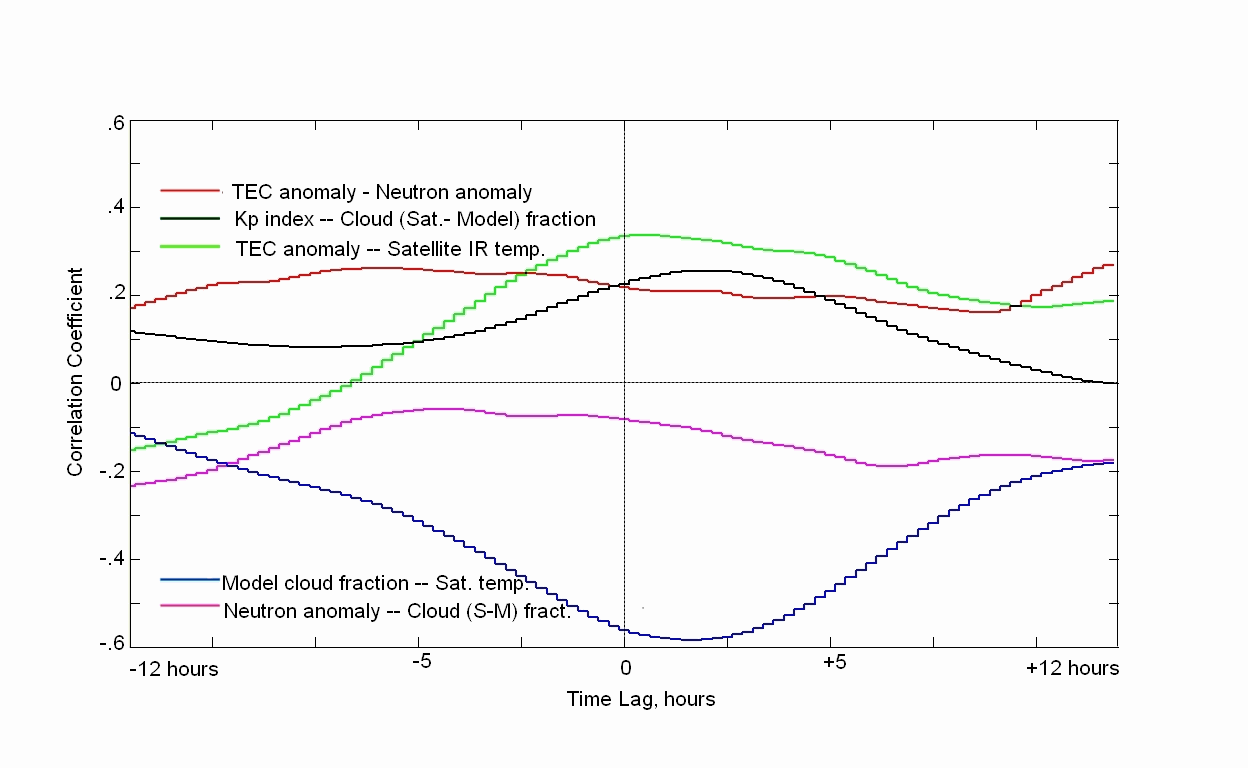} 
\caption{Time Series correlations with lags.}
\label{fig:cube2}
\end{center}
\end{figure}

In sum, the 5 time series offer some hints at possible connections which can be visualized through lagged cross-correlation analyses.  Figure 5 displays five of these correllelograms for the time series shown in figure 4. Unsurprisingly, the strongest correlation is seen between IR cloud temperature and forecast GFS cloud amount.  A peak correlation of -.57 is seen with a ~1.5 hour lag, indicating that the GFS forecast model does a fair job of predicting the presence and absence of cloudiness.  Higher IR temperatures (would be associated with clear skies) are, as expected, correlated strongly with GFS forecast of low cloud fractions (clearer skies).  The strength of this correlation and adequate model performance lends some credence to usage of anomalous measured-modeled cloud amount (bottom panel Fig. 4).  The TEC anomaly and Neutron monitored GCR flux show a modest positive correlation (in red), which is in agreement with previously reported results. The second strongest correlation (+.34, with very little lag) shows an association of higher than usual TEC values and clearer skies (higher IR Temp).  This correlation is opposite in sign to that predicted by most hypotheses.  The Granger-Causality test gives a confidence interval for this cross-correlation at 0.22 . The correlation of neutron monitored GCR flux and cloud anomaly also shows a correlation of opposite sign than has been hypothesized, though with less statistical significance.  During this event, decreases in GCR flux appear to correlate with positive cloud anomaly (presence of more clouds than expected). On the other hand, the Kp index and cloud anomaly are positively correlated (max. +.26 correlation at 2.5 hour lag).  While the correlations are indeed suggestive of possible relationships among a variety of measures, the analyses of a 10 day stretch for one location is insufficient to yield broad inferences about the connections in a globally-linked system.  Nevertheless, this example shows the promise of multi-modal comparisons of high resolution data streams:  with the unprecedented ionospheric imaging resolution promised by the inclusion of LWA and ETA data, and temporal and geographic extension of high-res analyses, the scientific rhubarb over cosmic/cloud connectivity can be made into a delicious pie.

\section{Conclusion}
We have laid out the principle elements for a comparison between solar, atmospheric, and cosmic ray observations conducted with LWA/ETA, satellite-based observations of cloud formations, and numerical weather models. Because of the ability of these low-frequency radio arrays to observe and track solar emissions, image the ionosphere with unprecedented resolution, and observe cosmic ray events, they are ideal for investigating a possible cosmic ray--cloud connection.  A sample case-study from a modest geomagnetic event in April 2010 suggests that this high-resolution approach has the potential to reveal unexpected connections among parameters in a complex and dynamic system.  The recent initiation of observations with LWA-1 as well coincident observations conducted by ETA will allow the program outlined here to begin to be carried out in the coming months. This effort will provide a great deal of insight into the possible existence of a causal relationship between cosmic rays and cloud formation rates by providing detailed observation over short time-scales of solar and cosmic ray events. Such observations will help establish if such an effect exists, and if no strong correlations are found, will help places limits on the role any potential such effect could have on climate forcing. 

\section*{Acknowledgments}
We thank Dr. Margaret Benoit for data reduction assistance, insightful comments and a careful reading of this manuscript.

GCR data was acquired from the Bartol Research Institute neutron monitor program, which is supported by the United States National Science Foundation under grants ANT-0739620 and ANT-0838839.
\bibliographystyle{unsrt}
\bibliography{reference}{}

\end{document}